\documentclass[aps,pra,reprint, amsmath, amssymb,superscriptaddress]{revtex4-1}
\usepackage[utf8]{inputenc}
\usepackage[T1]{fontenc}
\usepackage{graphicx}
\usepackage{grffile}
\usepackage{longtable}
\usepackage{wrapfig}
\usepackage{rotating}
\usepackage[normalem]{ulem}
\usepackage{amsmath}
\usepackage{textcomp}
\usepackage{amssymb}
\usepackage{capt-of}
\usepackage{hyperref}
\usepackage{tikz}
\usepackage{listings}
\usepackage{tikz}
\usepackage{pgfplots}
\pgfplotsset{compat=newest}
\pgfplotsset{every axis legend/.append style={%
cells={anchor=west}}
}
\usepgfplotslibrary{polar}
\usetikzlibrary{arrows}
\usetikzlibrary{graphs,decorations.pathreplacing,decorations.markings, positioning ,arrows,arrows.meta,decorations.pathmorphing,quotes,shapes.multipart,shapes.geometric,math,patterns}
\tikzset{>=stealth'}

\newcommand{\tikzeq}[1]{\begin{tikzpicture}[baseline={([yshift=-.5ex]current bounding box.center)}] #1 \end{tikzpicture}}
\usepackage{subfigure}
\usepackage{physics}
\usepackage{cleveref}
\usepackage[toc,page]{appendix}
\usepackage{amsthm}
\newtheorem{lemma}{Lemma}
\renewcommand{\vec}[1]{\boldsymbol{#1}}

\DeclareMathOperator{\NEG}{neg}
\DeclareMathOperator{\NS}{NL}

\date{\today}
\title{}
\begin{document}

\tikzset{%
   addarrow/.style = {postaction={decorate,decoration={markings,mark=at position #1 with {\arrow[]{Stealth}}}}},
   addarrow/.default = .72
   }
\title{Bell non-locality using tensor networks and sparse recovery}
\date{\today}

\author{I. S. Eliëns}
\affiliation{International Institute of Physics, Federal University of Rio Grande do Norte, 59078-970 Natal, Brazil}
\author{S. G. A. Brito}
\affiliation{International Institute of Physics, Federal University of Rio Grande do Norte, 59078-970 Natal, Brazil}
\author{R. Chaves}
\affiliation{International Institute of Physics, Federal University of Rio Grande do Norte, 59078-970 Natal, Brazil}
\affiliation{School of Science and Technology, Federal University of Rio Grande do Norte, 59078-970 Natal, Brazil}

\begin{abstract}
Bell's theorem, stating that quantum predictions are incompatible with a local hidden variable description, is a cornerstone of quantum theory and at the center of many quantum information processing protocols. Over the years, different perspectives on non-locality have been put forward as well as different ways to to detect non-locality and quantify it. Unfortunately and in spite of its relevance, as the complexity of the Bell scenario increases, deciding whether a given observed correlation is non-local becomes computationally intractable. Here, we propose to analyse a Bell scenario as a tensor network, a perspective permitting to test and quantify non-locality resorting to very efficient algorithms originating from compressed sensing and that offer a significant speedup in comparison with standard linear programming methods. Furthermore, it allows to prove that non-signalling correlations can be described by hidden variable models governed by a quasi-probability.
\end{abstract}

\maketitle{}

\section{Introduction}
Bell's theorem \cite{Bell1964} shows that quantum predictions are at odds with the physical intuition from classical physics. More precisely, that the correlations obtained by local measurements on distant but entangled systems cannot be reproduced by any local hidden variable model, the phenomenon generally known as Bell non-locality \cite{brunner2014bell}. Historically a topic in the foundations of quantum theory, with the establishment of quantum information science non-locality is now understood as a resource in a number of information processing applications ranging from randomness certification \cite{pironio2010random}, secure communication \cite{Ekert1991}, reduction in communication complexity \cite{buhrman2010nonlocality} and self-testing \cite{mayers2004self}. It is also at the core of the device-independent framework \cite{Acin2007} where information processing is achieved without the need of a precise knowledge of the internal physical mechanisms of the state preparation and measurement apparatuses.

A central problem in the study of non-locality is to decide whether a given observed correlation is non-local \cite{brunner2014bell} and furthermore quantify it \cite{de2014nonlocality}. The standard approach is that based on Bell inequalities, experimentally testable witnesses, the violation of which allows for the device-independent certification of the non-local nature of the correlations under test. The set of correlations compatible with a local hidden variable model is characterized by a convex set \cite{pitowsky1989quantum}, the non-trivial facets of which are precisely the Bell inequalities. However, the full characterization of Bell inequalities bounding a given scenario can only be achieved for the simplest cases \cite{brunner2014bell} and in practice one often has to rely on an incomplete set of inequalities \cite{Clauser1969,CGLMP,Collins_2004}. Alternatively, the non-local behaviour of a given correlation can be tested directly, resorting to linear programming (LP) \cite{brunner2014bell,2018_Brito_PRA_97,Baccari2017}. Notwithstanding, the LP approach also suffers from the curse of dimensionality, being of no use as the number of parties, measurements settings or measurement outcomes increase in the Bell scenario.

Motivated by these issues and the fact that new representations often lead to new insights \cite{pitowsky1989quantum,2011_Abramsky_NJP_13,acin2015combinatorial,fritz2012beyond,wood2015lesson,Causalbell2015}, our aim in this paper is to offer an alternative view on Bell non-locality, based on tensor networks \cite{2019_Orus_NRP_1} and sparse recovery \cite{2006_Donoho_ITIT_52,2007_Candes_PotICoMMA22,2012_Eldar,2016_Stern}. We also drew inspiration from category theory and its applications to quantum mechanics and probability theory which in some sense represents a formal counter part to the computation-focused tensor-network approach \cite{coecke2017picturing,coecke2016categorical,biamonte2011categorical,wetering2018quantum,fong2013causal,baez2014bayesian}. Tensor networks have become an important tool in condensed matter physics, and by itself constitutes a field in rapid development that branches out into topics as varied as quantum gravity and machine learning \cite{2019_Orus_NRP_1}. The current surge of progress can be traced back to the invention of the density matrix renormalization group (DMRG)  \cite{1992_White_PRL_69} and its reformulation in terms of matrix product states (MPS) \cite{2011_Schollwock_AoP_326}. The successful application of tensor networks, as well as machine learning models such as neural networks \cite{nielsen2015neural}, generally rest on a combination of two factors: (i) The computational problem at hand allows an efficient encoding in terms of the given model, (ii) there is an efficient manner to fix the free parameters of the model by means of an optimization problem. In the case of finding the ground state of one-dimensional systems the model is given the MPS ansatz and DMRG is the optimization algorithm. In the case of neural networks the algorithm is based on back propagation.

Here we show that the problem of determining Bell non-locality has a natural representation as a tensor network problem. It turns out that the optimization one has to perform is equivalent to the problem of basis pursuit known from the theory of compressed sensing. Compressed sensing refers to the idea that sparse signals can be reconstructed efficiently from a very limited number of observations (well below the Nyquist-Shannon limit) \cite{2006_Donoho_ITIT_52,2007_Candes_PotICoMMA22}. This allows to recover a signal from a small number of observed data points by using convex optimization for recovery. This has lead to many applications in the last decade  \cite{2012_Eldar,2016_Stern}. 

Based on the tensor network approach we show a number of results.
First we show that non-signalling correlations (including non-local correlations) can be mapped to hidden variable models governed by quasi-probabilities, that is, probabilities that sum up to one but are not necessarily  positive \cite{abramsky2014operational,appleby2017introducing}. Nicely, the negativity of this quasi-probability provides a natural way to quantify non-locality. Second, we provide an explicit singular value decomposition for the hidden variable model that introduces a natural basis to express the problem and points out a novel way to detect and quantify non-locality with tools originating from the field of compressed sensing \cite{2011_Becker_SJIS_4}. In fact, as we show, sparse recovery algorithms allow for a significant speed-up in the detection of non-locality in comparison with the standard linear programming approach.

\section{Bell scenario as a tensor network}
\label{sec:org105bdb2}
We will focus here on the standard bipartite Bell scenario in which Alice and Bob each locally  perform an experiment in spatially separated regions of spacetime. However, all our results generalize in a straightforward manner to more parties. Alice and Bob have the freedom to choose experimental settings, labeled \(x\) and \(y\) respectively, and obtain outcomes indexed by labels  \(a\) and
\(b\) with some probability. We will assume that \(x,y\) can both take values  \(1,\ldots,m\)
while  \(a,b\) take one of the values \(0 ,\ldots,n-1\). The setup is fully described by a conditional probability \(P(ab\vert xy)\), i.e. the
probability to obtain outcome \(a\) and \(b\) given the inputs \(x\) and \(y\). We will call $\vec{P}$ a behavior. 

In order to be consistent with the laws of special relativity, the behavior \(\vec{P}\) must obey the no-signalling conditions
\begin{equation}
\label{eq:4}
\begin{split}
 \sum_a P(ab|xy)  &= \sum_a P(ab'|xy)\qquad \text{for all $b,b'$},\\
 \sum_b P(ab|xy)  &= \sum_b P(a'b|xy)\qquad \text{for all $a,a'$}.
\end{split}
\end{equation}

In a quantum description, according to Born's rule the probability distribution in a Bell scenario should be given by
\begin{equation}
P(ab \vert xy)= \mathrm{Tr} \left[ \left(M_a^x \otimes M_b^y \right) \rho \right],
\end{equation}
where $\rho$ is the density matrix describing the quantum state shared between Alice and Bob and $M_a^x$ and $M_b^y$ are POVM operators describing their measurements. Clearly, quantum correlations are non-signalling, however, there are non-signalling correlations of a post-quantum nature \cite{popescu1994quantum}.

If moreover the experimental outcomes can be explained within the assumptions of local realism, the conditional probability allows a
decomposition 
\begin{equation}
P(ab|xy) = \sum_{\lambda} P(a|x\lambda)P(b|y\lambda)
p(\lambda) .  
\end{equation}
It is well known that we can replace the local conditional probabilities (e.g. \(P(a|x \lambda)\) for Alice) by a deterministic process mapping each \(x\) to some \(a\) (i.e. a function) and the local
variable simply determines the probability for the combination of deterministic processes at Alice's and Bob's side. In other words, \(\lambda\) can be taken to be the combination of two sequences \((a_1\ldots a_m)\) and \((b_1\ldots b_m)\) that prescribe the outputs \(a_x\) and \(b_y\) for each of the inputs \(x\) and \(y\) and we can write
\begin{equation}
\label{eq:NS}
P(ab|xy) = \sum_{a_1\ldots a_m}\sum_{b_1\ldots b_m}\delta_{a a_x}\delta_{b b_x} q_{a_1\ldots a_m b_1\ldots b_m},
\end{equation}
where \(q_{a_1\ldots a_m b_1\ldots b_m}\) is the corresponding
probability. Remarkably, as we will show next, \emph{any no-signalling} conditional
probability allows such a decomposition with \(\vec{q}\) a \emph{quasi probability}, i.e. 
 \(q_{a_1\ldots a_m b_1
\ldots b_m}\) can take negative values but still sums to 1. This was noted in
\cite{2011_Abramsky_NJP_13} but in slightly different form and totally
different language. We will arrive at this observation independently from a reasoning rooted in tensor network theory. 
Furthermore, the decomposition in Eq. \eqref{eq:NS} gives a new approach to testing
Bell non-locality: One can now search the space of all \(\vec{q}\) compatible with \(\vec{P}\) for an element with only non-negative coefficients, i.e. a proper probability (see Sec. \ref{sec:recovery}). 

Let us establish some conventions and notations. Any
object with multiple indices such as a conditional probability
\(P(ab|xy)\) or a (quasi) probability \(q_{a_1\ldots a_m}\) will be viewed
as a tensor. We will use mixed notations for indices---upper, lower,
as function argument---without distinction. For clarity, we will not
use the Einstein summation convention and keep summations explicit.  If any or all of the indices are suppressed we will use
bold face such as \(\vec{P}\) and \(\vec{q}\). 
The graphical depiction of a tensor as a box with a line
for each index is often useful. For example, \(\vec{P}\) we depict as
\begin{equation}
\label{eq:1}
 P(ab|xy) = 
\tikzeq{
\node (x) at (.5,0) {$x$};
\node (y) at (1.5,0) {$y$};
\node (a) at (.5,-2) {$a$};
\node (b) at (1.5,-2)  {$b$};
\node[draw, rectangle, minimum width = 1.5cm] (P) at (1,-1)  {$P$};
\draw [addarrow ]  (x) -- (x |- P.north);
\draw  [addarrow ]  (a |- P.south) -- (a); 
\draw [addarrow ]  (y) -- (y |- P.north);
\draw  [addarrow ]  (b |- P.south) -- (b);
},
\end{equation}
where we used an arrow on the lines to distinguish input from output
indices. Connecting lines between tensors implies summation over the
corresponding index, also called \emph{contraction}. This way one can create tensor networks
representing a bunch of tensors with a certain pattern of contractions.

The decomposition in Eq. \eqref{eq:NS} of any no-signalling \(P(ab|xy)\) can be shown using a typical tensor-network tool, namely the singular value decomposition (SVD). Recall that any matrix \(\vec{M}\) allows an SVD \(\vec{M} = \vec{U} \vec{S} \vec{V}^{\dag}\) where \(\vec{U},\vec{V}\)
are unitary (orthogonal if \(\vec{M}\) is real) and \(\vec{S}\) is quasi diagonal. Furthermore, we will make use of the following lemma.
\begin{lemma}
\label{lemma1}
A matrix $M_{ax}$ has constant column sums $\sum_a M_{ax} = \mathcal{C}$ (independent of $x$) iff it can be decomposed as $M_{ax} = \sum_{a_1\ldots a_m} \delta_{a a_x} C_{a_1\ldots a_m}$ with $\sum_{a_1\ldots a_m}C_{a_1\ldots a_m} = \mathcal{C}$. If all $M_{ax}\geq 0$ then all $C_{a_1\ldots a_m} \geq 0$. (Here we assume that $x$ takes values $1,\ldots ,m$.)  
\end{lemma}
\emph{Proof:} To show the \emph{if} statement, suppose that $M_{ax} = \sum_{a_1\ldots a_m} \delta_{a a_x} C_{a_1\ldots a_m}$. Summing over $a$ gives $\sum_a M_{ax} = \sum_{a_1\ldots a_m} C_{a_1\ldots a_m} = \mathcal{C}$. Clearly $M_{ax} \geq 0$ if  $C_{a_1\ldots a_m} \geq 0$ and the statement follows.

To show the \emph{only if} statement, let us start with the case $M_{ax} \geq 0$. We do induction on the column sum. For $\mathcal{C} = 0$, the only option is $M_{ax} = 0$ for all $a,x$. Suppose we proved the statement for $\mathcal{C'} \leq \mathcal{C}$ and we are given $M_{ax} \geq 0$ with column sums $\mathcal{C}$. Pick the coefficients $a_x$ of the smallest non-zero elements  of each column $x$ of $\vec{M}$ and let $\lambda$ be the smallest value of all these elements. Then matrix $\vec{M}'$ with coefficients $M'_{ax} = M_{ax} -\lambda \delta_{aa_x}$ and $M'_{a'x'} = M_{a'x'}$ for $(a,x)\neq (a',x')$, has all coefficients non-negative and column sums $\mathcal{C}' = \mathcal{C} - \lambda \leq \mathcal{C}$. Hence, by the induction step we can write $M'_{ax} = \sum_{a'_1\ldots a'_m} \delta_{aa'_x}C'_{a'_1\ldots a'_m}$ with $C'_{a_1\ldots a_m} \geq 0$. Adding back $\lambda \delta_{aa_x}$ gives the required decomposition for $\vec{M}$. This establishes the lemma for this case.

Next, note that the matrices with a single 1 and a single -1 in one of the columns and otherwise zeros can be constructed as $N_{ax} = \delta_{aa_x} - \delta_{aa'_x}$ where $\vec{a} = (a_1,\ldots ,a_m)$ and $\vec{a}' = (a'_1\ldots,a'_m)$ differ only for the label $x$ corresponding to the column with $a_x \neq a'_x$  the corresponding row indices of the non-zero coefficients. We can convert any matrix with constant column sums into any other matrix with the same constant column sums by adding a superposition of such matrices $\vec{N}$. Combined with the case $M_{ax}\geq 0$ we have shown that the matrices $\delta_{aa_x}$ form an over-complete set that generates all matrices with constant column sum and the lemma follows.  $\qed$

The conditional probability $P(a|x)$ can be viewed as a stochastic matrix with constant column sums $1$ and non-negative coefficients. The lemma states that this can be decomposed as a superposition of matrices with a single 1 in each column with non-negative coefficients, as in the following simple example:
\begin{equation}
  \left[\begin{matrix}1/8 & 3/8\\7/8 & 5/8 \end{matrix} \right] =
\frac{1}{8}
  \left[\begin{matrix}1 & 0\\0 & 1 \end{matrix} \right]+
\frac{3}{8}
  \left[\begin{matrix}0 & 1\\1 & 0 \end{matrix} \right]+
  \frac{4}{8}\left[\begin{matrix}0 & 0\\1 & 1 \end{matrix} \right] 
\end{equation}
Such a decomposition is not necessarily unique. The white noise probability $P(a|x) = 1/n$ can for example easily be seen to allow different decompositions. If we would allow negative coefficients there is of course an even larger ambiguity in choosing the coefficients. For a conditional quasi probability $Q(a|x)$, a matrix with column sums equal to 1 but some negative elements, we will always find some negative expansion coefficients in the super position, such as in the following simple example:
\begin{equation}
  \left[\begin{matrix}-1/8 & 3/8\\9/8 & 5/8 \end{matrix} \right] =
 -\frac{1}{8}\left[\begin{matrix}1 & 1\\ 0 & 0\end{matrix} \right]+
\frac{4}{8}
  \left[\begin{matrix}0 & 1\\1 & 0 \end{matrix} \right]+
  \frac{5}{8}\left[\begin{matrix}0 & 0\\1 & 1 \end{matrix} \right] 
\end{equation}
The non-negative case of the lemma essentially states that stochastic matrices form a polytope with corners given by the matrices with coefficients $\delta_{aa_x}$ which simply encode functions or deterministic processes. As such it can be viewed as a generalization of the Birkhoff-von Neumann theorem. 

Let us now consider the bipartite $P(ab|xy)$ that satisfies the no-signalling property and see how the lemma implies the decomposition in terms of a quasi probability for the hidden variable model.  We can interpret  \(\vec{P}\) as a matrix by
grouping Alice's and Bob's indices and write the SVD as \(P(ab|xy) =\sum_\lambda
A_{ax}^{\lambda}\Lambda_{\lambda}B_{by}^{\lambda}\). From the no-signalling
property it is clear that \(\sum_a A_{ax}^{\lambda}\) is independent of
\(x\) for each \(\lambda\) and similar for \(\sum_b B_{by}^{\lambda}\) is
independent of \(y\). Then, applying the
lemma for fixed \(\lambda\) we find that \(A_{ax}^{\lambda} = \sum_{a_1\ldots a_m}
\delta_{aa_x} \mathcal{A}^{\lambda}_{a_1\ldots a_m}\) and \(B_{ax}^{\lambda} = \sum_{a_1\ldots a_m}
\delta_{bb_x} \mathcal{B}^{\lambda}_{b_1\ldots b_m}\) for some
\(\vec{\mathcal{A}}^{\lambda}\) and \(\vec{\mathcal{B}}^{\lambda}\). Hence we have found a
decomposition of the form in Eq. \eqref{eq:NS} with
\(q_{a_1\ldots a_m b_1\ldots b_m} = \sum_{\lambda}
\mathcal{A}^{\lambda}_{a_1\ldots a_m}\mathcal{B}^{\lambda}_{b_1\ldots b_m}\Lambda_{\lambda}\) which is real
and summing over all indices gives 1. However, in general the
coefficients of \(\vec{q}\) can be negative hence it is a quasi probability.

To give a concrete example, let us put $n = m = 2$ and consider the Popescu-Rohrlich (PR) box \cite{popescu1994quantum}
\begin{equation}
    P(ab|xy) = \frac{1}{2} \delta_{a \oplus b, xy}
\end{equation}
(for the fully binary case we  use $x,y \in \{ 0,1\}$ and $a\oplus b = a+b \mod 2$). We can find
\begin{equation}
    q_{a_0a_1 b_0b_1} = \frac{1}{16}
    \begin{bmatrix}
        3 &3 & -1 & -1 \\
        3& -1 & 3 & -1 \\
        -1 & 3 & -1 & 3 \\
        -1 & -1 & 3 & 3 
    \end{bmatrix}_{a_0a_1,b_0b_1}.
\end{equation}
Doing the summation we recover $P(ab|xy) = \sum_{a_0a_1b_0b_1}\delta_{aa_x}\delta_{bb_y} q_{a_0a_1b_0b_1}$.

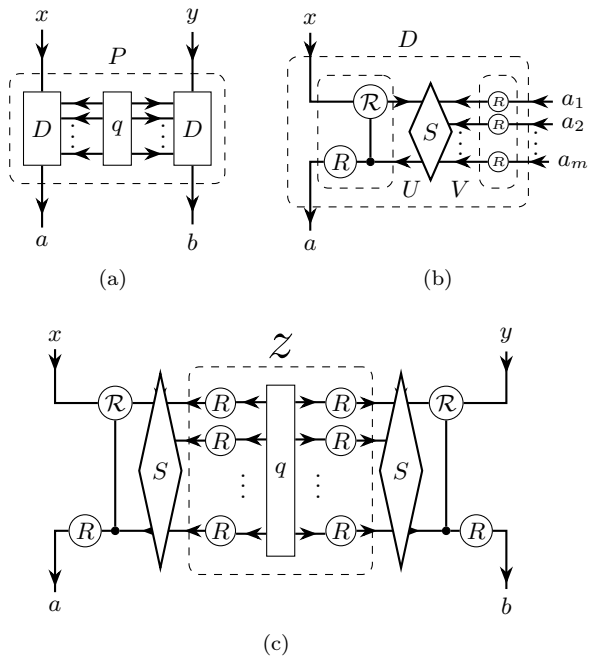
\begin{figure}[t]
\hspace{.5cm}
\subfigure[]{
\begin{tikzpicture}
    \node (x) at (0,0.5) {$x$};
    \node (y) at (2,0.5) {$y$};
    \node (a) at (0,-2.5) {$a$};
    \node (b) at (2,-2.5) {$b$};
    \node [draw, rectangle, minimum height=3em] (D1) at (0,-1) {$D$};
    \node [draw, rectangle, minimum height=3em] (D2) at (2,-1) {$D$};
    \node [draw, rectangle, minimum height=3em] (q) at (1,-1) {$q$};
  
    \draw[thick, addarrow] (q.120) -- (q.120 -| D1.east);
    \draw[thick, addarrow] (q.145) -- (q.145 -| D1.east);
    \draw[thick, addarrow] (q.240) -- (q.240 -| D1.east);

    \draw[thick, addarrow] (q.60) -- (q.60 -| D2.west);
    \draw[thick, addarrow] (q.35) -- (q.35 -| D2.west);
    \draw[thick, addarrow] (q.300) -- (q.300 -| D2.west);

     \draw[thick, addarrow =  .5] (x) -- (D1);
     \draw[thick, addarrow = .8] (D1) -- (a);

     \draw[thick, addarrow = .5] (y) -- (D2);
     \draw[thick, addarrow = .8] (D2) -- (b);

     \node [right] at (.25,-1 ) { $\vdots$};
     \node [right] at (1.45,-1 ) { $\vdots$};

     \node [draw, rectangle, rounded corners, dashed, minimum height = 4.5 em, minimum width = 8.8 em] (P) at (q) {};
     \node [above] at (P.north) {$P$};
 \end{tikzpicture}
 }\hfill
\subfigure[]{
\begin{tikzpicture}
    \node (x) at (0,0.5) {$x$};
    \node (a) at (0,-2.5) {$a$};
    
    \node[draw, circle,inner sep = 1pt] (R) at ( .4,-1.4) {$R$};
    \node[draw, circle,inner sep = 1pt] (CR) at (.8,-.6) {$\mathcal{R}$};
    \node[draw, circle, inner sep=1pt, fill=black] (copy) at (.8,-1.4) {};
      
    \draw [thick, addarrow =.93] (R.west) -| (a);
    \draw [thick, addarrow =.23] (x) |- (CR.west);
    \draw [thick] (R.east) -- (copy);
    \draw [thick] (copy) -- (CR);
    
    \node (S) at ( 1.6,-1) {};
    \draw [thick, addarrow = .31] (CR) -| (S.center);
    \draw [thick, addarrow = .78] (S.center) |- (copy);

    \node [circle, draw, inner sep = .5pt] at ( 2.5,-.6) (R1) {\tiny $R$};
    \node [circle, draw, inner sep = .5pt] at ( 2.5,-.9) (R2) {\tiny $R$};
    \node [circle, draw, inner sep = .5pt] at ( 2.5,-1.4) (Rm) {\tiny $R$};
    \node at (2.0,-1.05) {$\vdots$}; 

    \node  at ( 3.5,-.6) (a1) {$a_1$};
    \node at ( 3.5,-.9) (a2) {$a_2$};
    \node at ( 3.5,-1.4) (am) {$a_m$};
    \node at (3.0,-1.05) {$\vdots$};

    \draw [thick, addarrow=.5] (a1) -- (R1);
    \draw [thick, addarrow=.5] ( a2) -- (R2);
    \draw [thick, addarrow=.5] (am) -- (Rm);
    
    \draw [thick, addarrow=.6]  (R1) -- ++ (-.9,0);
    \draw [thick, addarrow=.6]  (R2) -- ++ (-.9,0) ;
    \draw [thick, addarrow=.6]  (Rm) -- ++ (-.9,0) ;

    \node[draw, thick, diamond, minimum height = 4 em, inner sep = 1pt, fill = white] at (S) {$S$};
    \node[ draw, rectangle, rounded corners, dashed, minimum width = 3.2cm, minimum height = 2.0cm]  (D) at (1.3,-1)  {};
    \node[above] at (D.north) {$D$};

    \node[ draw, rectangle, rounded corners, dashed, minimum width = 1cm, minimum height = 1.5cm]  (U) at (0.6,-1)  {};
    \node[right] at (U.south east) {$U$};
\node[ draw, rectangle, rounded corners, dashed, minimum width = .5cm, minimum height = 1.5cm]  (V) at (2.5,-1)  {};
    \node[left] at (V.south west) {$V$};

 \end{tikzpicture}

}
\subfigure[]{
\begin{tikzpicture}
    \node (x) at (0,0.8) {$x$};
    \node (y) at (6,0.8) {$y$};
    \node (a) at (0,-2.8) {$a$};
    \node (b) at (6,-2.8) {$b$};

    \node [draw, rectangle, minimum height=7em] (q) at (3,-1) {$q$};
  
    \node [circle, draw, inner sep = .7pt] at ( 2.2,-.1) (Ra1) {$R$};
    \node [circle, draw, inner sep = .7pt] at ( 2.2,-0.6) (Ra2) {$R$};
    \node [circle, draw, inner sep = .7pt] at ( 2.2,-1.8) (Ram) {$R$};
    \node at (2.5,-1.1) {$\vdots$}; 
    
    \draw[thick, addarrow = .8] (q.102) -- (q.102 -| Ra1.east);
    \draw[thick, addarrow] (q.115) -- (q.115 -| Ra2.east);
    \draw[thick, addarrow] (q.257) -- (q.257 -| Ram.east);

    \draw [thick, addarrow=.4]  (Ra1) -- ++ (-.9,0);
    \draw [thick, addarrow=.4]  (Ra2) -- ++ (-.9,0) ;
    \draw [thick, addarrow=.4]  (Ram) -- ++ (-.9,0) ;
    
    \node[draw, circle,inner sep = 1pt] (R) at ( .4,-1.8) {$R$};
    \node[draw, circle,inner sep = 1pt] (CR) at (.8,-0.1) {$\mathcal{R}$};
    \node[draw, circle, inner sep=1pt, fill=black] (copy) at (.8,-1.8) {};
      
    \draw [thick, addarrow =.93] (R.west) -| (a);
    \draw [thick, addarrow =.23] (x) |- (CR.west);
    \draw [thick] (R.east) -- (copy);
    \draw [thick] (copy) -- (CR);
    
    \node (S) at ( 1.4,-1) {};
    \draw [thick, addarrow = .31] (CR) -| (S.center);
    \draw [thick, addarrow = .78] (S.center) |- (copy);
       
     \node[draw, thick, diamond, minimum height = 8 em, inner sep = 1pt, fill = white] at (S) {$S$};
     
     \node [circle, draw, inner sep = .7pt] at ( 3.8,-.1) (Rb1) { $R$};
    \node [circle, draw, inner sep = .7pt] at ( 3.8,-0.6) (Rb2) { $R$};
    \node [circle, draw, inner sep = .7pt] at ( 3.8,-1.8) (Rbm) { $R$};
    \node at (3.5,-1.1) {$\vdots$}; 
    
    \draw[thick, addarrow] (q.78) -- (q.78 -| Rb1.west);
    \draw[thick, addarrow] (q.65) -- (q.65 -| Rb2.west);
    \draw[thick, addarrow] (q.283) -- (q.283 -| Rbm.west);

    \draw [thick, addarrow=.4]  (Rb1) -- ++ (.9,0);
    \draw [thick, addarrow=.4]  (Rb2) -- ++ (.9,0) ;
    \draw [thick, addarrow=.4]  (Rbm) -- ++ (.9,0) ;

    \node[draw, circle,inner sep = 1pt] (Rb) at ( 5.6,-1.8) {$R$};
    \node[draw, circle,inner sep = 1pt] (CRb) at (5.2,-0.1) {$\mathcal{R}$};
    \node[draw, circle, inner sep=1pt, fill=black] (copyb) at (5.2,-1.8) {};
      
    \draw [thick, addarrow =.93] (Rb.east) -| (b);
    \draw [thick, addarrow =.23] (y) |- (CRb.east);
    \draw [thick] (Rb.west) -- (copyb);
    \draw [thick] (copyb) -- (CRb);
    
    \node (Sb) at ( 4.6,-1) {};
    \draw [thick, addarrow = .31] (CRb) -| (Sb.center);
    \draw [thick, addarrow = .78] (Sb.center) |- (copyb);
       
     \node[draw, thick, diamond, minimum height = 8 em, inner sep = 1pt, fill = white] at (Sb) {$S$};
     
     \node [draw, rectangle, rounded corners, dashed, minimum height = 8.5 em, minimum width = 7.5 em] (z) at (q) {};
     \node [above] at (z.north) {\huge{$z$}};
     
 \end{tikzpicture}
 }
\caption[]{\label{fig:D} Tensor-network decomposition of bipartite no-signalling conditional probability $P(ab|xy)$. (a)  The no-signalling condition on $\vec{P}$ is equivalent to the decomposition $\vec{P} = (\vec{D}\otimes \vec{D}) \vec{q}$ with $\vec{q}$ a quasi probability and $\vec{D}$ the deterministic tensor. (b) The singular value decomposition (SVD) of deterministic tensor $\vec{D}$. Writing $\vec{D} = \vec{U} \vec{S} \vec{V}^T$ we find that $\vec{U}$ and $\vec{V}$ are expressed in terms of the rotation $\vec{R}$. (c) The full decomposition of $\vec{P}$. Indicated is the vector $\vec{z} = \vec{R}^{2m}\vec{q}$, i.e. the hidden-variable quasi probability in the basis of singular vectors of $\vec{D} \otimes \vec{D}$ (correlation basis). It is natural to compare $\vec{q}$ with the behavior $\vec{P}$ in this basis as discussed in the text and is used to obtain $\vec{q}$ via sparse recovery.}
\end{figure}

We can refine the decomposition \eqref{eq:NS} further. Let us introduce the deterministic tensor \(\vec{D}\)
defined by the 
coefficients \(D_{axa_1 \ldots a_m} = \delta_{aa_x}\). Viewing \(\vec{D}\) as a matrix we can apply the
singular value decomposition (SVD)
\begin{align}
\label{eq:D}
D_{ax a_1\ldots a_m} &=\delta_{a a_x} \\
&=\sum_{a'x'}\sum_{a_1'\ldots a_m'}U_{ax a'x'} S_{a'x' a'_1\ldots a'_m} V_{a_1\ldots a_m a'_1\ldots a'_m}.\notag
\end{align}
Remarkably, we can find exact analytic results for this SVD.
By a direct computation one can check that the following definitions
are consistent with this decomposition
\begin{align}
\label{eq:5a}
 U_{axa'x'} &= R_{aa'} \mathcal{R}^{a'}_{xx'},\\
\label{eq:5b}
S_{axa_1\ldots a_m} &= \sqrt{n^{m-1}}\left[\sqrt{m}\delta_{0a}\delta_{1x} + 1- \delta_{0a} \right]\notag\\
&\qquad \times\delta_{aa_x}\prod_{y\neq x} \delta_{0a_y},\\
\label{eq:5c}
V_{a_1\ldots a_m,a'_1\ldots a'_m} &= R_{a_1a'_1}\ldots R_{a_ma'_m},
\end{align}
where
\begin{equation}
\label{eq:6}
 R_{ab} = \begin{cases}
1/\sqrt{n} \qquad&\text{for $b=0$},\\
1/\sqrt{b(b+1)}\qquad&\text{for $a<b\neq 0$},\\
-b/\sqrt{b(b+1)}\qquad&\text{for $a=b\neq 0$},\\
0 \qquad&\text{for $a>b\neq 0$},
\end{cases}
\end{equation}
and 
\begin{equation}
\label{eq:Rcal}
 \mathcal{R}^{a}_{xy} = \delta_{0a}R_{xy} + (1 - \delta_{0a}) \delta_{xy}.
\end{equation}
Note that \(\vec{R}\) defines a rotation to a basis of which the first
vector labeled by \(b=0\) has all coefficients equal and is thus the normalized joint
eigenvector with all coefficients equal to 1 of any stochastic matrix,
while all other basis  vectors \(b>0\) have
column sum zero. The tensor \(\vec{\mathcal{R}}\) is like the controlled version
of \(\vec{R}\) which only implements the rotation when the control equals
\(a=0\) and which for \(a \neq 0\) just acts as the identity, similar in
spirit to the well-known CNOT gate. In the definition of \(R_{xy}\) we
have to replace \(n\) by \(m\) when compared to \(R_{ab}\). Note that for the case \(n=2\) the
matrix \(\vec{R}\) is simply the Hadamard gate 
\begin{equation}
\vec{R} = \frac{1}{\sqrt{2}} \begin{bmatrix} 1&1\\1&-1\end{bmatrix}    
\end{equation}
 while for $n = 3$
 \begin{equation}
     \vec{R} = \begin{bmatrix}
     1/\sqrt{3} & 1/\sqrt{2} & 1/\sqrt{6}\\ 
     1/\sqrt{3} & -1/\sqrt{2} & 1/\sqrt{6}\\1/\sqrt{3} & 0 & -2/\sqrt{6}
     \end{bmatrix}.
 \end{equation}
Note also that for a binary
probability \(p_a\) with \(a = 0,1\) the vector \(\vec{z} =\vec{R} \vec{p}\)
has only a single non-trivial coefficient \(z_1 = \sum_a(-1)^a
p_a/\sqrt{2}\) which up to normalization
equals the imbalance or expectation value \(\langle a \rangle = \sum_a (-1)^a
p_a\). We
will call the basis defined by the columns of \(\vec{R}\) the
\emph{correlation basis} since rotating a binary probability \(\vec{p}\) by the
corresponding transformation is analogues to switching to a
description in terms of the correlators (expectation values) rather than the probability
itself. This becomes clearer still if we consider a probability with
more indices \(p_{a_1\ldots a_m}\) and apply \(\vec{R}\) as we will
discuss in the next section.  The 
rotation \(\vec{R}\) straightforwardly generalizes to the case \(n>2\) in
which case the coefficients \(z_{a'}\) with \(a'>0\) together are a
generalized correlation-like description of \(p_a\) that is
mathematically equivalent.

Lemma \ref{lemma1} and the explicit SVD of \(\vec{D}\) [Eqs. \eqref{eq:D} to \eqref{eq:Rcal}] represent the main technical
results of this paper. The first implies the decomposition in
Eq. \eqref{eq:NS} which, although known, we provide with a new derivation
and utility based on tensor networks. Refining the decomposition using the SVD of \(\vec{D}\) suggests to attack the problem of Bell
non-locality in the basis of singular vectors of \(\vec{D} \otimes \vec{D}\) for the hidden variable of Alice and Bob. Noting the tensor product structure of the matrix \(\vec{V} = \vec{R} \otimes \ldots \otimes \vec{R}\) it follows that this is exactly the correlation basis which we have just described. 

\section{The correlation basis}
\label{sec:org8fc7eeb}
The problem of Bell non-locality is exactly equivalent to
the marginal problem: given \(P(ab|xy)\) can we find the probability
\(p_{a_1\ldots a_m b_1 \ldots b_m}\) that reproduces all correct marginals?
Given the  probability \(\vec{p}\) we can construct the marginals such as \(P(a|x)\) by summing over all indices
except \(a_x\) and \(P(ab|xy)\) by summing over all indices except \(a_x\)
and \(b_y\) and so on. Using the tensor network structure in the Bell non-locality
problem we have layed out before we 
can see exactly how this equivalence is manifest in the basis of
singular vectors of \(\vec{D}\otimes \vec{D}\) in the bipartite
scenario (which we call the correlation basis). Specifying \(P(ab|xy)\)
fixes the coefficients of basis elements with non-zero singular
values. The kernel of \(\vec{D}\otimes \vec{D}\) then determines a
subspace of quasi probabilities \(\vec{q}\) which are compatible with
\(\vec{P}\). Whether or not this subspace contains a proper probability
\(\vec{p}\) is then a subsequent problem which can be treated with
algorithms from the theory of compressed sensing. This will be 
discussed in the next section.

Let us make a general definition: For a general (quasi) probability \(q_{c_1\ldots c_l}\) we can use the
rotation \(\vec{R}\) to define
\begin{equation}
\label{eq:2}
 z_{c_1'\ldots c_l'} = \sum_{c_1\ldots c_l} R_{c_1c'_1}\ldots R_{c_lc'_l} q_{c_1\ldots c_l},
\end{equation}
and we refer to the standard basis in the \(c_z'\) indices as the
\emph{correlation basis}.
In the bipartite scenario the \(c_z\) indices are split into two groups
\(a_x\) and \(b_y\). But because \(\vec{z}\) is defined by acting on \(\vec{q}\)
by a tensor product operator \(\vec{V} = \vec{R} \otimes \ldots \otimes
\vec{R}\), the definition generalizes to multipartite scenarios or when
Alice and Bob have different numbers of inputs (or even outputs).

\begin{figure}[t]
\centering
\begin{minipage}[b]{.3\linewidth}
\centering
\includegraphics[width = \linewidth]{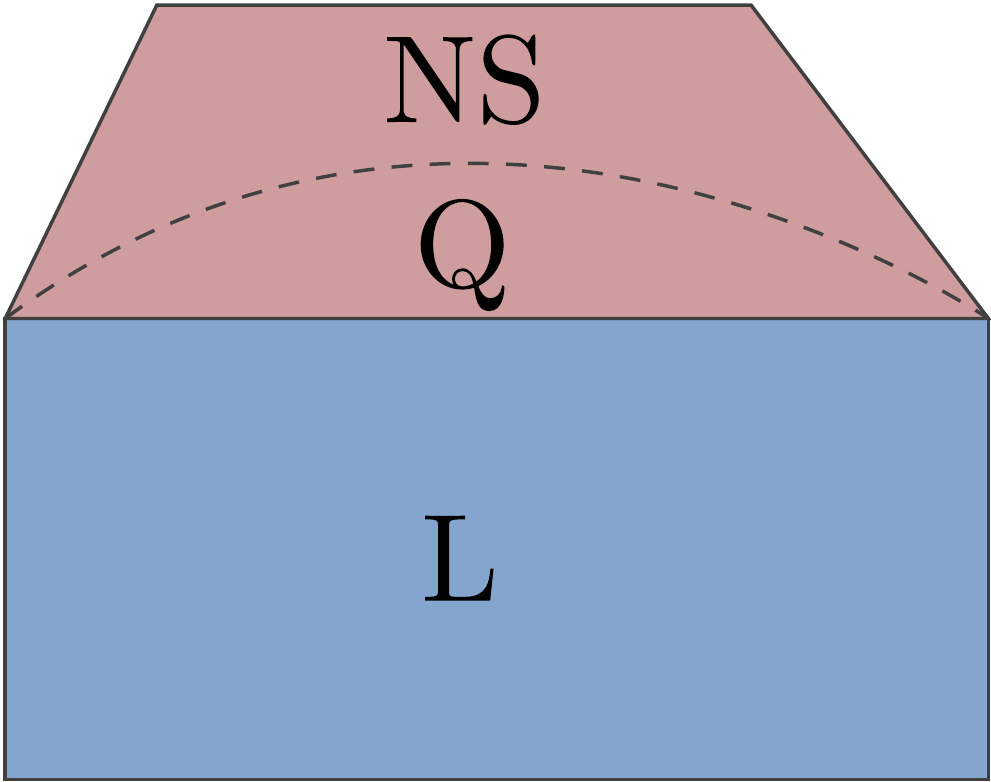}\\
(a)
\end{minipage}%
\hspace{-.7cm}
\begin{minipage}[t!]{.7\linewidth}
\centering
\includegraphics[width = \linewidth]{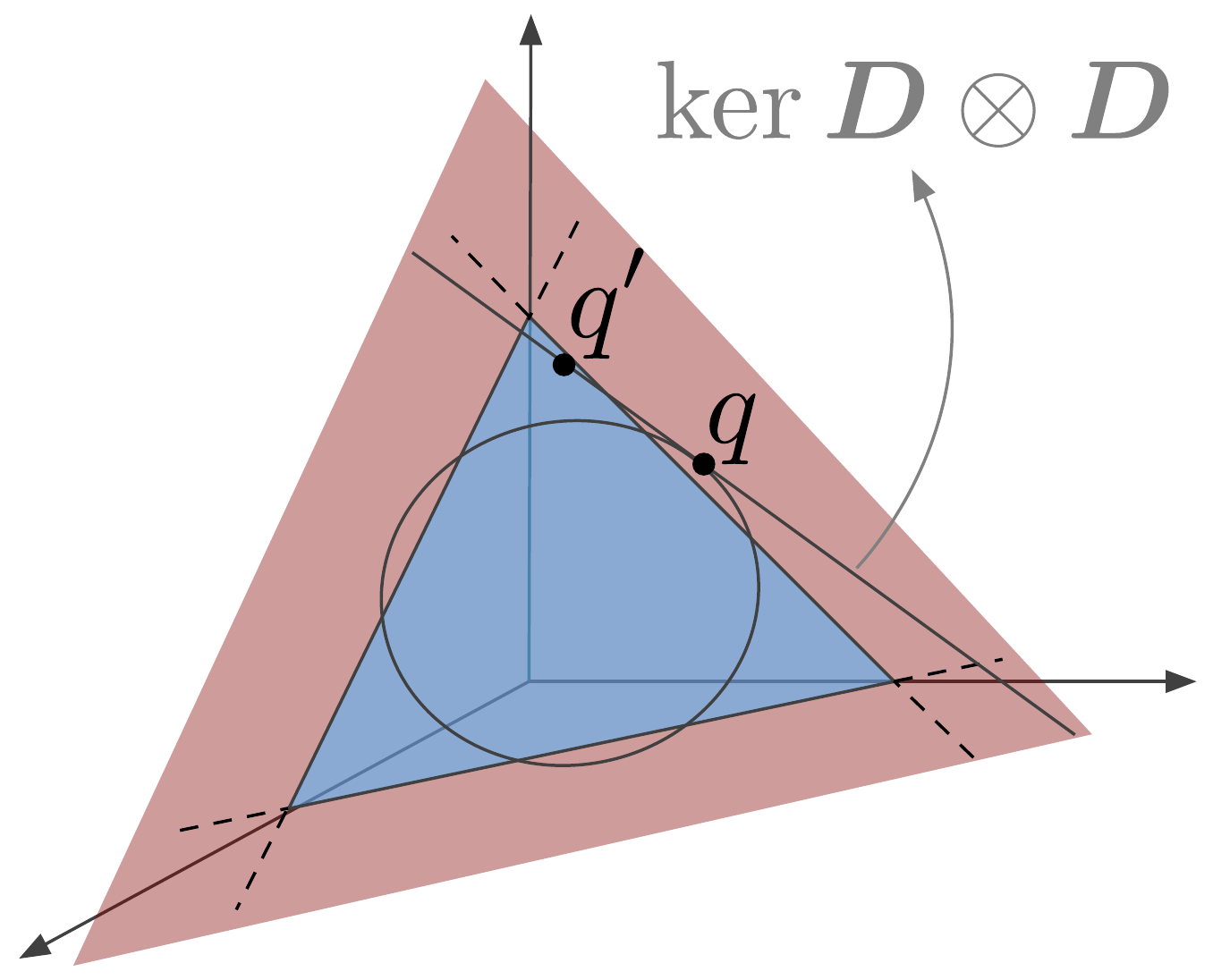}\\
(b)
\end{minipage}
\caption[]{\label{fig:simplex} Conceptual illustration of the local versus nonlocal probabilities. (a) In the space of conditional probabilities $\vec{P}$ the no-signalling condition defines a polytope (NS). Strictly included in NS there is the polytope of local correlations (L). We show that NS exactly corresponds to those $\vec{P}$ which allow a hidden variable model defined by a quasi probability $\vec{q}$, i.e. solutions to the equation $(\vec{D}\otimes \vec{D})\vec{q} = \vec{P}$. The local polytope L exactly corresponds to those $\vec{Q}$ for which $\vec{q}$ can be chosen to non-negative, i.e. a probability. The quantum set Q is defined  as $\vec{P}$ that can be obtained from a quantum mechanical model, $P(ab|xy) = \Tr[\vec{M}^x_a\otimes \vec{M}^y_b \vec{\rho}]$. It is not a polytope and satisfies the strict inclusions $\mathrm{L} \subset \mathrm{Q} \subset \mathrm{NS}$. (b) Illustration of the hyperplane of quasi probabilities $\vec{q}$  (red) with the probability simplex indicated (blue). It is shown that a solution $\vec{q}$ for the hidden variable quasi probability that minimizes \(\ell_2\)-norm can exist even when $\vec{P}$ is local if $\vec{q}' =\vec{q} + \vec{k}$  where $\vec{k} \in \ker \vec{D}\otimes \vec{D}$.}
\end{figure}

Given a no-signalling \(P(ab|xy)\), we can obtain unambiguous marginals
\(P(a|x)\) and \(P(b|y)\). Given the decomposition \eqref{eq:NS} it is
straightforward to see that these are obtained by summing over all
indices except \(a_x\) and/or \(b_y\). By the fact that \(R_{a0} =
1/\sqrt{n}\) we can easily see that
\begin{align}
\label{eq:z}
 z_{0\ldots00\ldots 0} &= n^{-m},\\
 z_{0\ldots a'_x\ldots 00\ldots 0} &= n^{\frac{1}{2}-m} \sum_{a} R_{a a'_x} P(a|x),\\
 z_{0\ldots00\ldots  b'_y\ldots 0} &= n^{\frac{1}{2}-m} \sum_{b} R_{b b'_y} P(b|y),\\
 z_{0\ldots a_x'\ldots 00\ldots  b'_y\ldots 0} &= n^{1-m} \sum_{b}R_{a a'_x} R_{b b'_y} P(ab|xy),
\label{eq:zend}
\end{align}
so these elements are fixed by the given \(P(ab|xy)\). Equivalently, one look at the decomposition of \(\vec{P}\) in terms of \(\vec{z}\) as 
\begin{equation}
\label{eq:9}
 \vec{P} = (\vec{U} \vec{S} \otimes \vec{U}\vec{S})\vec{z}
\end{equation}
and formally invert this by taking the inverse only of the non-zero
singular values. The corresponding coefficients of \(\vec{z}\) are
exactly the ones fixed by the marginals that follow from the
problem. Putting all other coefficients of \(\vec{z}\) to zero
corresponds to the \(\vec{q}\) obtained by applying the
Moore-Penrose pseudo inverse \cite{penrose1955generalized} of \(\vec{D}\otimes \vec{D}\) on
\(\vec{P}\). This gives the solution of the linear equation
\((\vec{D}\otimes \vec{D} ) \vec{q} = \vec{P}\) of minimal
\(\ell_2\)-norm. While a viable quasi probability consistent with
\(\vec{P}\), there is no guarantee that this \(\vec{q}\) is a probability if
\(\vec{P}\) is local since adding any element \(\vec{k}\) of the kernel of
\(\vec{D}\otimes\vec{D}\) to \(\vec{q}\) gives a \(\vec{q}' = \vec{q} +
\vec{k}\) that is still a quasi probability and reproduces
\(\vec{P}\). It may happen that \(\vec{q}\) with minimal \(\ell_2\)-norm lies outside
the probability simplex but some \(\vec{q}'\) lies inside the
probability simplex. The clearest example is given by the deterministic cases that form the corners of the local hidden variable polytope. Let us consider $n=m=2$ and $P(00|xy) = 1$ and all other $P(ab|xy)=0$. The correct hidden variable probability that generates this behavior is
\begin{equation}
    \vec{q}' = 
\begin{bmatrix}
 1 & 0 & 0 & 0 \\
 0 & 0 & 0 & 0 \\
 0 & 0 & 0 & 0 \\
 0 & 0 & 0 & 0 .
\end{bmatrix}
\end{equation}
but the minimal $\ell_2$-norm solution is in fact
\begin{equation}
\vec{q} = \frac{1}{16}\begin{bmatrix}
 9 & 3 & 3 & -3 \\
 3 & 1 & 1 & -1 \\
 3 & 1 & 1 & -1 \\
 -3 & -1 & -1 & 1 .
\end{bmatrix}
\end{equation}
This is illustrated in Fig. \ref{fig:simplex}. 

One way to guarantee that a solution \(\vec{q}\) is a proper probability is to make sure it minimizes an \(\ell_p\)-norm with \(0\leq p \leq 1\). Algorithms to solve this problem will be discussed in the
next section.

Bell inequalities are the traditional way to test Bell
non-locality. We can offer a new perspective on Bell inequalities in terms of the correlation basis. Clearly, any local conditional probability \(\vec{P}\) has by definition a proper probability  \(\vec{q}= \vec{p}\) for the hidden variable. For \(\vec{q}\) to be a
probability it simply has have non-negative coefficients, which in
terms of \(\vec{z}\) means \(\vec{\mathcal{V}} \vec{z} \vec{\geq} \vec{0}\)
(the inequality is interpreted as element wise). Here
\(\vec{\mathcal{V}} = \vec{R}^{\otimes 2m}\). One can use
quantifier elimination methods such as Fourier-Motzkin elimination to derive
the domains for a restricted set of coefficients of \(\vec{z}\) which can satisfy these inequalities treating the other coefficients as free variables that can take any required value. The Bell setup amounts to fixing certain coefficients of \(\vec{z}\) while any value for the other coefficients leads to a consistent \(\vec{q}\). Hence if the reduced set of inequalities is satisfied by the fixed
coefficients of \(\vec{z}\) we can find a true probability consistent with \(\vec{P}\). Hence, these inequalities are the Bell inequalities. Alternatively, one can proceed as usual, write down the deterministic strategies (the extremal points of the local polytope) in terms of the correlation basis and use standard convex optimization algorithms to obtain Bell inequalities.

To illustrate this point let us give the CHSH inequality   \cite{Clauser1969} in the correlation basis. This corresponds to $n=m=2$ in terms of which the non-trivial Bell inequality in terms of  $z_{a_0a_1 b_0 b_1}$ reads
\begin{equation}
   | z_{1010} + z_{0110} + z_{1001} - z_{0101} |\leq \frac{1}{2}
.
\end{equation}
The transformation to the correlation basis may also be useful in attacking more general contextuality scenarios \cite{Spekkens2005}. 

\section{Recovering the hidden variable and quantifying nonlocality}
\label{sec:recovery}

We will now consider the following problem: given a conditional
probability \(\vec{P}\) for a bipartite Bell scenario, determine whether
it corresponds to a local hidden variable model. As a further
refinement of the problem one may be interested in quantifying by some
distance measure the degree of non-locality in case \(\vec{P}\) is found
to be non-local. This problem is also considered in Ref. \cite{2018_Brito_PRA_97}.  Here we will show how techniques from compressed
sensing can be used to solve this problem efficiently. 

In compressed sensing one is interested in solving the linear equation
\(\vec{A} \vec{x} = \vec{b}\) and find the solution \(\vec{x}\) that is as
sparse as possible. Formally, the most sparse solution minimizes the
\(\ell_0\)-norm but instead minimizing the \(\ell_1\)-norm is known to
give a good approximation and it has the benefit of being convex and
amenable to the techniques of convex optimization.

In our case we can also focus attention on the \(\ell_1\)-norm, and we
consider the optimization problem
\begin{equation*}
\label{eq:10}
\begin{array}{rl}
\text{minimize} & ||\vec{q}||_1\\
\text{subject to} & (\vec{D} \otimes \vec{D})\vec{q} = \vec{P}
\end{array}
\end{equation*}
This problem is known as basis pursuit in the computer science
literature. There are several classes of algorithms equipped to solve
this. A well-known way is to map the problem to a linear program
(LP). Although slightly differently formulated, this is very similar
to the approach detailed in Ref. \cite{2018_Brito_PRA_97}. The problem
with LPs is however that, while accurate, for large dimensions they
become computationally expansive. For instance, the dimension of the hidden-variable
space grows exponentially with the number of inputs as \(n^{2m}\).

We will follow a different route here: Based on the tensor-network decomposition of \(\vec{P}\) we can formulate the problem such that it precisely fits the most efficient version of NESTA (a shorthand for Nesterov's algorithm), a class of
algorithms introduced in \cite{2011_Becker_SJIS_4} to tackle exactly the basis pursuit problem. 

Recall that fixing the no-signalling conditional probability \(P(ab|xy)\) is equivalent to fixing certain
coefficients of the vector \(\vec{z}\) [Eqs. \eqref{eq:z} to \eqref{eq:zend}]. Let us introduce the projector
\(\vec{\Pi}\) that projects onto these coefficients and let
us denote the vector of these coefficients by \(\vec{z}_0\). Then the
problem can be reformulated as
\begin{equation*}
\begin{array}{rl}
\text{minimize} & ||\vec{\mathcal{V}}\vec{z} ||_1\\
\text{subject to} & \vec{\Pi} \vec{z} = \vec{z}_0
\end{array}
\end{equation*}
where \(\vec{\mathcal{V}} = \vec{V} \otimes \vec{V} = \vec{R}^{\otimes
2m}\). We have used the tensor-network decomposition of \(\vec{P}\) and the NCON function in Matlab \cite{2014_Pfeifer_ACPP_}  to compute the vector
\(\vec{z}_0\) from \(\vec{P}\). Subsequently we have fed this into the NESTA package
available online \cite{2011_Becker_SJIS_4,NESTAonline} to obtain \(\vec{q} =
\vec{\mathcal{V}}\vec{z}\). 
The main goal here is efficiency. If \(\vec{q}\) is found to be non-negative
we have obtained a probability by construction and the original
\(\vec{P}\) is local. 

\begin{figure}[t]
\subfigure[]{
\begin{tikzpicture}[]
\begin{axis}[x = 3.8mm, y = 3.35mm,  ylabel = {$10^5\, \mathrm{sec}$},  xmin = {1.46}, xmax = {20.54}, ymax = {10.880077842484923}, xlabel = {$n$}, unbounded coords=jump,scaled x ticks = false,xlabel style = {font = {\fontsize{11 pt}{14.3 pt}\selectfont}, color = {rgb,1:red,0.00000000;green,0.00000000;blue,0.00000000}, draw opacity = 1.0, rotate = 0.0},xmajorgrids = true,xtick = {2.0,3.0,4.0,5.0,6.0,7.0,8.0,9.0,10.0,11.0,12.0,13.0,14.0,15.0,16.0,17.0,18.0,19.0,20.0},xticklabels = {$2$,$3$,$4$,$5$,$6$,$7$,$8$,$9$,$10$,$11$,$12$,$13$,$14$,$15$,$16$,$17$,$18$,$19$,$20$},xtick align = inside,xticklabel style = {font = {\fontsize{8 pt}{10.4 pt}\selectfont}, color = {rgb,1:red,0.00000000;green,0.00000000;blue,0.00000000}, draw opacity = 1.0, rotate = 0.0},x grid style = {color = {rgb,1:red,0.00000000;green,0.00000000;blue,0.00000000},
draw opacity = 0.1,
line width = 0.5,
solid},axis x line* = left,x axis line style = {color = {rgb,1:red,0.00000000;green,0.00000000;blue,0.00000000},
draw opacity = 1.0,
line width = 1,
solid},scaled y ticks = false,ylabel style = {font = {\fontsize{11 pt}{14.3 pt}\selectfont}, color = {rgb,1:red,0.00000000;green,0.00000000;blue,0.00000000}, draw opacity = 1.0, rotate = 0.0},ymajorgrids = true,ytick = {0.0,2.5,5.0,7.5,10.0},yticklabels = {$0.0$,$2.5$,$5.0$,$7.5$,$10.0$},ytick align = inside,yticklabel style = {font = {\fontsize{8 pt}{10.4 pt}\selectfont}, color = {rgb,1:red,0.00000000;green,0.00000000;blue,0.00000000}, draw opacity = 1.0, rotate = 0.0},y grid style = {color = {rgb,1:red,0.00000000;green,0.00000000;blue,0.00000000},
draw opacity = 0.1,
line width = 0.5,
solid},axis y line* = left,y axis line style = {color = {rgb,1:red,0.00000000;green,0.00000000;blue,0.00000000},
draw opacity = 1.0,
line width = 1,
solid},    xshift = 0.0mm,
    yshift = 0.0mm,
    axis background/.style={fill={rgb,1:red,1.00000000;green,1.00000000;blue,1.00000000}}
,title style = {font = {\fontsize{14 pt}{18.2 pt}\selectfont}, color = {rgb,1:red,0.00000000;green,0.00000000;blue,0.00000000}, draw opacity = 1.0, rotate = 0.0},legend style = {color = {rgb,1:red,0.00000000;green,0.00000000;blue,0.00000000},
draw opacity = 1.0,
line width = 1,
solid,fill = {rgb,1:red,1.00000000;green,1.00000000;blue,1.00000000},font = {\fontsize{8 pt}{10.4 pt}\selectfont}},colorbar style={title=}, ymin = {-0.31689225364926954}, width = {152.4mm}]\addplot+ [color = {rgb,1:red,0.51372549;green,0.00000000;blue,0.00000000},
draw opacity = 0.39,
line width = 1,
solid,mark = none,
mark size = 2.0,
mark options = {
    color = {rgb,1:red,0.00000000;green,0.00000000;blue,0.00000000}, draw opacity = 1.0,
    fill = {rgb,1:red,0.51372549;green,0.00000000;blue,0.00000000}, fill opacity = 0.39,
    line width = 1,
    rotate = 0,
    solid
},forget plot]coordinates {
(2.0, 0.0003908976112318502)
(3.0, 0.001173724376502661)
(4.0, 0.0033146723402675193)
(5.0, 0.009169937370664412)
(6.0, 0.0251834646288889)
(7.0, 0.06897875893557824)
(8.0, 0.1887542320390723)
(9.0, 0.5163274150550978)
(10.0, 1.4122052392114024)
(11.0, 3.8623358281240274)
(12.0, 10.563182462405653)
};
\addplot+ [color = {rgb,1:red,0.33333333;green,0.71372549;blue,1.00000000},
draw opacity = 0.65,
line width = 1,
dashed,mark = none,
mark size = 2.0,
mark options = {
    color = {rgb,1:red,0.00000000;green,0.00000000;blue,0.00000000}, draw opacity = 1.0,
    fill = {rgb,1:red,0.33333333;green,0.71372549;blue,1.00000000}, fill opacity = 0.65,
    line width = 1,
    rotate = 0,
    solid
},forget plot]coordinates {
(2.0, 0.016952591010908482)
(3.0, 0.027990231424384733)
(4.0, 0.041187284560451905)
(5.0, 0.05696621963642515)
(6.0, 0.07583215808228454)
(7.0, 0.09838904368509838)
(8.0, 0.12535897627332343)
(9.0, 0.15760532785843173)
(10.0, 0.19616038123681506)
(11.0, 0.24225837582968138)
(12.0, 0.29737501863727156)
(13.0, 0.36327472514748943)
(14.0, 0.4420671024933529)
(15.0, 0.5362744830201598)
(16.0, 0.6489126701732786)
(17.0, 0.7835874815754578)
(18.0, 0.9446101798683291)
(19.0, 1.1371354865352692)
(20.0, 1.367326596858009)
};
\addplot+ [color = {rgb,1:red,0.33333333;green,0.71372549;blue,1.00000000},
draw opacity = 0.65,
line width = 1,
solid,mark = none,
mark size = 2.0,
mark options = {
    color = {rgb,1:red,0.00000000;green,0.00000000;blue,0.00000000}, draw opacity = 1.0,
    fill = {rgb,1:red,0.33333333;green,0.71372549;blue,1.00000000}, fill opacity = 0.65,
    line width = 1,
    rotate = 0,
    solid
},forget plot]coordinates {
(2.0, 0.0023840328537230613)
(4.0, 0.006421744362806726)
(6.0, 0.013260204749427101)
(8.0, 0.024842146736656658)
(10.0, 0.044457874928634206)
(12.0, 0.07768000580494179)
(14.0, 0.1339465878640216)
(16.0, 0.22924234924665324)
(18.0, 0.3906397863250266)
(20.0, 0.6639901674702589)
};
\addplot+ [color = {rgb,1:red,0.33333333;green,0.71372549;blue,1.00000000},
draw opacity = 0.65,
line width = 1,
solid,mark = none,
mark size = 2.0,
mark options = {
    color = {rgb,1:red,0.00000000;green,0.00000000;blue,0.00000000}, draw opacity = 1.0,
    fill = {rgb,1:red,0.33333333;green,0.71372549;blue,1.00000000}, fill opacity = 0.65,
    line width = 1,
    rotate = 0,
    solid
},forget plot]coordinates {
(3.0, 0.02414915928664992)
(5.0, 0.05340104043585179)
(7.0, 0.1010334512875956)
(9.0, 0.17859586691402532)
(11.0, 0.3048949185520297)
(13.0, 0.510554455014736)
(15.0, 0.8454409358672019)
(17.0, 1.3907545983731273)
(19.0, 2.2787181547360666)
};
\addplot+ [color = {rgb,1:red,0.00000000;green,0.00000000;blue,0.00000000},
draw opacity = 0.0,
line width = 1,
solid,mark = diamond*,
mark size = 4.5,
mark options = {
    color = {rgb,1:red,0.00000000;green,0.00000000;blue,0.00000000}, draw opacity = 1.0,
    fill = {rgb,1:red,0.33333333;green,0.71372549;blue,1.00000000}, fill opacity = 0.65,
    line width = 1,
    rotate = 0,
    solid
}]coordinates {
(2.0, 9.614110000000001e-5)
(3.0, 0.00201949874)
(4.0, 0.0008661741)
(5.0, 0.00841060753)
(6.0, 0.00394522854)
(7.0, 0.044626394320000005)
(8.0, 0.01160895415)
(9.0, 0.08862724399)
(10.0, 0.03237322973)
(11.0, 0.28630154468)
(12.0, 0.06608626782)
(13.0, 0.48975442069)
(14.0, 0.13069333474)
(15.0, 0.85379763217)
(16.0, 0.24343027573999998)
(17.0, 1.50992844374)
(18.0, 0.40053071057)
(19.0, 2.22087242316)
(20.0, 0.6568375759599999)
};
\addlegendentry{NESTA}
\addplot+ [color = {rgb,1:red,0.00000000;green,0.00000000;blue,0.00000000},
draw opacity = 0.0,
line width = 1,
solid,mark = *,
mark size = 3.5,
mark options = {
    color = {rgb,1:red,0.00000000;green,0.00000000;blue,0.00000000}, draw opacity = 1.0,
    fill = {rgb,1:red,0.51372549;green,0.00000000;blue,0.00000000}, fill opacity = 0.39,
    line width = 1,
    rotate = 0,
    solid
}]coordinates {
(2.0, 3.1264300000000002e-6)
(3.0, 1.718306e-5)
(4.0, 0.00010934961999999999)
(5.0, 0.00069046417)
(6.0, 0.00454893065)
(7.0, 0.02084149104)
(8.0, 0.10376494846)
(9.0, 0.39003998879)
(10.0, 1.49846818478)
(11.0, 3.8717650563399997)
(12.0, 10.55626578794)
};
\addlegendentry{LP}
\end{axis}

\end{tikzpicture}
}
\subfigure[]{
\input{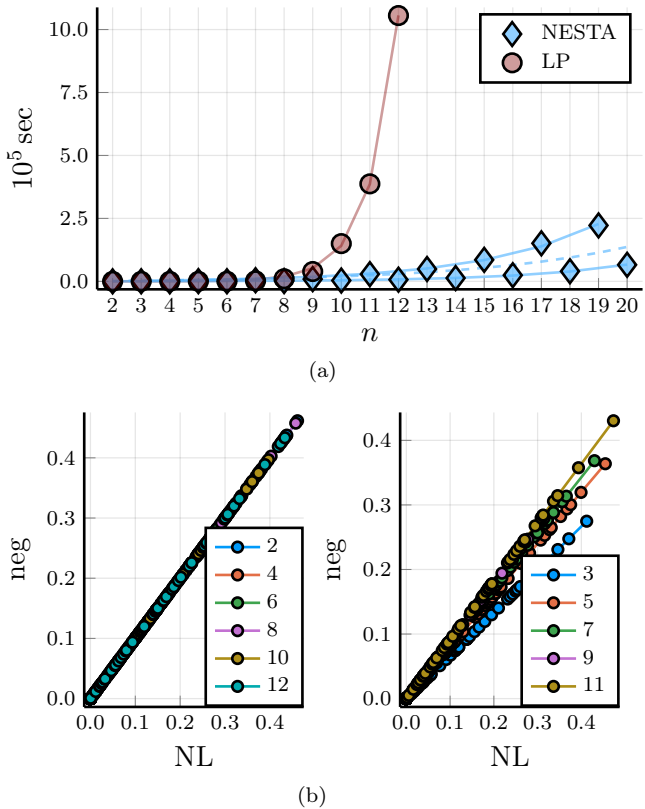}
}
\caption[]{\label{fig:plots} 
(a) Computation times for Bell $22n$ (bipartite, two inputs, $n$ outputs). We computed $q$ via the NESTA algorithm which gives access to $\NEG(\vec{P})$ and we computed the $\ell_1$ distance $\NS(\vec{P})$ via linear programming (LP). Non-zero values of $\NEG(\vec{P})$ and $\NS(\vec{P})$ correspond to non-locality of $\vec{P}$. We see that the NESTA computation scales much more favourably with increasing dimension. The sampled data here is 100 points of a random convex combination of the form $\vec{P} = c_0\vec{P}_{\mathrm{wn}} + c_1 \vec{P}_{\mathrm{ld}} + c_2 \vec{P}_{\mathrm{pr}}$ where $P_{\mathrm{wn}}(ab|xy) = 1/n^2$ is white noise, $P_{\mathrm{ld}}(ab|xy) = \delta_{a0}\delta_{b0}$  is local deterministic  and $P_{\mathrm{pr}}(ab|xy) = 1/n$ if $b-a = xy \mod 2$ and zero otherwise is a generalized PR box, i.e. a corner of the no-signalling set. Solid lines are fits with $f(n) = a [\exp(b n)-1]$. For LP we find $b \approx 1.006$. For NESTA we find a distinguished difference between even and odd cases, which we fit separately. This gives an exponent $b\approx 0.25$. Fitting the even and odd cases together (dashed blue line) gives $b \approx 0.18$. (b) We compare $\NS(\vec{P})$ with $\NEG(\vec{q})$ showing that all cases which are deemed local by the LP method are indeed local according to the NESTA based method. The relation is linear in all cases, but while in the cases $n$ is even we find equality $\NEG(\vec{q}) = \NS(\vec{P})$, the cases with $n$ odd have a non-trivial constant of proportionality. 
}\end{figure}

In Fig. \ref{fig:plots} we have
compared the LP based method from \cite{2018_Brito_PRA_97} implemented in Mathematica with our
current method. We see that at least in the case of varying number of outputs we really do get a very significant improvement in the computation time for larger cases using the NESTA based approach. To test the correctness we have also compared $\NS(\vec{P})$ with $\NEG(\vec{P})$ and see that the relation is linear. Here $\NS(\vec{P})$ is the $\ell_1$ distance of $\vec{P}$ to the local polytope \cite{2018_Brito_PRA_97}, while $\NEG(\vec{P})$ is the minimal negativity $\NEG(\vec{q}) = \sum_{a_1\ldots a_m} \sum_{b_1\ldots b_m} \max( -q_{a_1\ldots a_mb_1\ldots b_m},0)$ for all $\vec{q}$ compatible with $\vec{P}$ and can also be understood as a measure of how non-local a given distribution $\vec{P}$ is. Since we also have $\NEG(\vec{q}) = \frac{1}{2}||\vec{q}||_1 - 1$, minimizing the negativity is indeed equivalent to minimizing the $\ell_1$ norm. Remarkably, for even $n$ we find exact equality $\NEG(\vec{P}) = \NS(\vec{P})$ and better performance of the NESTA algorithm while for odd $n$ we find a non-trivial constant of proportionality and also longer computation times. An exact diagnosis of why this is the case is postponed to future research. One remark we can make however is that the longer computation time seems to be caused by the fact that more iterations are needed in NESTA to reach the stop criterion. A big part of the computational cost of NESTA comes from the matrix multiplication by $\vec{\mathcal{V}}$ and $\vec{\mathcal{V}}^T$. We leveraged the fact that $\vec{\mathcal{V}} = \vec{R}^{\otimes2m}$ is a tensor product operator by using a multiplication routine that does not construct the full matrix $\vec{\mathcal{V}}$ \cite{kronmult}.

\section{Conclusions}
\label{sec:org2e8df1a}
Bell non-locality is a cornerstone of quantum theory and central resource in a variety of quantum information processing protocols. For that aim, a basic step is to decide whether a given observed correlation is non-local or not. Given its importance, over the years a few approaches been developed to tackle to the problem \cite{brunner2014bell,Fonseca2015,montina2016can,Baccari2017,2018_Brito_PRA_97,Duarte2018,Canabarro2019}, however, limited in practice to relatively simple Bell scenarios, with a small number of distant parties and measurement settings and outcomes. Most of such approaches are based on a geometric view of Bell scenarios, more precisely the fact that local correlations form a polytope and that to detect non-locality means to find ways of certifying that a given correlation lies outside this local set. However, different perspectives such as sheaf-theoretic \cite{2011_Abramsky_NJP_13}, graph theoretical \cite{acin2015combinatorial}, causal  \cite{wood2015lesson,Causalbell2015} or category theoretic \cite{wolfe2019quantifying,rosset2019characterizing,schmid2019typeindependent} are also possible. Often a different  perspective leads to new insights and computational methods. 

Here we propose a new perspective to understand and quantify Bell non-locality, based on a tensor network description. With that we showed that non-signalling correlations can be described by hidden variable models governed by quasi-probabilities, the negativity of which offers a natural way to quantify non-locality. By refining our description via a singular value decomposition we obtained a natural basis to attack the problem of quantifying non-locality and that can be computationally implemented via extremely efficient sparse recovery algorithms. To show its relevance we compared the NESTA algorithm used in compressed sensing with the standard linear programming tool used in the study of non-locality and showed a significant speed-up in computational time as function of the complexity of the Bell scenario.

We believe this perspective on Bell's theorem opens a few venues for future research. Local correlations are equivalent to hidden variable models governed by probabilities while general non-signalling correlations imply quasi-probabilities. What are the restrictions imposed by Born's rule (the quantum mechanical rule) to such quasi-probabilities? There is an important research program trying to recover the quantum limitations on correlations \cite{popescu1994quantum,Navascues2007}, however, to our knowledge, the intersection of this program with this quasi-probability description of non-locality has not been considered so far (see however recent results on general connections between quantum theory and quasi probabilities such as \cite{zhu2016quasiprobability,appleby2017introducing,wetering2018quantum}). A potential application of the sparse recovery method is to combine it with the machine-learning approach that has been recently proposed \cite{Canabarro2019}. There, neural networks are used in supervised learning algorithms that create a machine model of the local set. The bottleneck of the method is exactly the fact that the standard linear programming approach is used to generate the input data, consisting of a set of correlations and their respective degree of non-locality. Can the NESTA algorithm \cite{2011_Becker_SJIS_4} provide a more scalable solution to this machine learning approach? Finally, the tensor network description can easily be extended to more complex Bell scenarios consisting of several independent sources \cite{branciard2010characterizing} and leading to non-convex sets of correlations \cite{PhysRevLett.116.010402,PhysRevLett.116.010403}. Can generalizations of sparse recovery algorithms adapted to deal with non-linear constraints \cite{ohlsson2013quadratic} provide a new way to deal with such complicated Bell scenarios? Can tensor network ideas be further leveraged to menage computational cost if the network of causal relations  increases in size in a certain way? We hope our results might motivate further research along these directions.

\begin{acknowledgments}
We acknowledge the John Templeton Foundation via the Grant Q-CAUSAL No. 61084, the Serrapilheira Institute (Grant No. Serra-1708-15763), the Brazilian National Council for Scientific and Technological Development (CNPq) via the National Institute for Science and Technology on Quantum Information (INCT-IQ) and Grants No. 307172/2017-1 and No. 406574/2018-9, the Brazilian agencies MCTIC and MEC.
\end{acknowledgments}

\bibliographystyle{unsrturl} 
\bibliography{../references.bib}

\begin{thebibliography}{10}

\bibitem{Bell1964}
J.~S. Bell.
\newblock On the {E}instein--{P}odolsky--{R}osen paradox.
\newblock {\em Physics}, 1:195--200, 1964.

\bibitem{brunner2014bell}
Nicolas Brunner, Daniel Cavalcanti, Stefano Pironio, Valerio Scarani, and
  Stephanie Wehner.
\newblock Bell nonlocality.
\newblock {\em Reviews of Modern Physics}, 86(2):419, 2014.

\bibitem{pironio2010random}
Stefano Pironio, Antonio Ac{\'\i}n, Serge Massar, A~Boyer de~La~Giroday,
  Dzmitry~N Matsukevich, Peter Maunz, Steven Olmschenk, David Hayes, Le~Luo,
  T~Andrew Manning, et~al.
\newblock Random numbers certified by bell’s theorem.
\newblock {\em Nature}, 464(7291):1021, 2010.

\bibitem{Ekert1991}
Artur~K. Ekert.
\newblock Quantum cryptography based on bell's theorem.
\newblock {\em Phys. Rev. Lett.}, 67:661--663, Aug 1991.
\newblock URL: \url{https://link.aps.org/doi/10.1103/PhysRevLett.67.661}, \href
  {http://dx.doi.org/10.1103/PhysRevLett.67.661}
  {\path{doi:10.1103/PhysRevLett.67.661}}.

\bibitem{buhrman2010nonlocality}
Harry Buhrman, Richard Cleve, Serge Massar, and Ronald De~Wolf.
\newblock Nonlocality and communication complexity.
\newblock {\em Reviews of modern physics}, 82(1):665, 2010.

\bibitem{mayers2004self}
Dominic Mayers and Andrew Yao.
\newblock Self testing quantum apparatus.
\newblock {\em Quantum Information \& Computation}, 4(4):273--286, 2004.

\bibitem{Acin2007}
Antonio Ac\'{\i}n, Nicolas Brunner, Nicolas Gisin, Serge Massar, Stefano
  Pironio, and Valerio Scarani.
\newblock Device-independent security of quantum cryptography against
  collective attacks.
\newblock {\em Phys. Rev. Lett.}, 98:230501, Jun 2007.
\newblock URL: \url{https://link.aps.org/doi/10.1103/PhysRevLett.98.230501},
  \href {http://dx.doi.org/10.1103/PhysRevLett.98.230501}
  {\path{doi:10.1103/PhysRevLett.98.230501}}.

\bibitem{de2014nonlocality}
Julio~I De~Vicente.
\newblock On nonlocality as a resource theory and nonlocality measures.
\newblock {\em Journal of Physics A: Mathematical and Theoretical},
  47(42):424017, 2014.

\bibitem{pitowsky1989quantum}
Itamar Pitowsky.
\newblock {\em Quantum probability-quantum logic}.
\newblock Springer, 1989.

\bibitem{Clauser1969}
John~F. Clauser, Michael~A. Horne, Abner Shimony, and Richard~A. Holt.
\newblock Proposed experiment to test local hidden-variable theories.
\newblock {\em Phys. Rev. Lett.}, 23:880--884, Oct 1969.
\newblock URL: \url{https://link.aps.org/doi/10.1103/PhysRevLett.23.880}, \href
  {http://dx.doi.org/10.1103/PhysRevLett.23.880}
  {\path{doi:10.1103/PhysRevLett.23.880}}.

\bibitem{CGLMP}
Daniel Collins, Nicolas Gisin, Noah Linden, Serge Massar, and Sandu Popescu.
\newblock Bell inequalities for arbitrarily high-dimensional systems.
\newblock {\em Phys. Rev. Lett.}, 88:040404, Jan 2002.
\newblock URL: \url{https://link.aps.org/doi/10.1103/PhysRevLett.88.040404},
  \href {http://dx.doi.org/10.1103/PhysRevLett.88.040404}
  {\path{doi:10.1103/PhysRevLett.88.040404}}.

\bibitem{Collins_2004}
Daniel Collins and Nicolas Gisin.
\newblock A relevant two qubit bell inequality inequivalent to the {CHSH}
  inequality.
\newblock {\em Journal of Physics A: Mathematical and General},
  37(5):1775--1787, jan 2004.
\newblock URL: \url{https://doi.org/10.1088%2F0305-4470%2F37%2F5%2F021}, \href
  {http://dx.doi.org/10.1088/0305-4470/37/5/021}
  {\path{doi:10.1088/0305-4470/37/5/021}}.

\bibitem{2018_Brito_PRA_97}
S.~G.~A. Brito, B.~Amaral, and R.~Chaves.
\newblock Quantifying {{Bell}} nonlocality with the trace distance.
\newblock {\em Physical Review A}, 97(2):022111, February 2018.
\newblock \href {http://dx.doi.org/10.1103/PhysRevA.97.022111}
  {\path{doi:10.1103/PhysRevA.97.022111}}.

\bibitem{Baccari2017}
F.~Baccari, D.~Cavalcanti, P.~Wittek, and A.~Ac\'{\i}n.
\newblock Efficient device-independent entanglement detection for multipartite
  systems.
\newblock {\em Phys. Rev. X}, 7:021042, Jun 2017.
\newblock URL: \url{https://link.aps.org/doi/10.1103/PhysRevX.7.021042}, \href
  {http://dx.doi.org/10.1103/PhysRevX.7.021042}
  {\path{doi:10.1103/PhysRevX.7.021042}}.

\bibitem{2011_Abramsky_NJP_13}
Samson Abramsky and Adam Brandenburger.
\newblock The sheaf-theoretic structure of non-locality and contextuality.
\newblock {\em New Journal of Physics}, 13(11):113036, 2011.
\newblock \href {http://dx.doi.org/10.1088/1367-2630/13/11/113036}
  {\path{doi:10.1088/1367-2630/13/11/113036}}.

\bibitem{acin2015combinatorial}
Antonio Ac{\'\i}n, Tobias Fritz, Anthony Leverrier, and Ana~Bel{\'e}n Sainz.
\newblock A combinatorial approach to nonlocality and contextuality.
\newblock {\em Communications in Mathematical Physics}, 334(2):533--628, 2015.

\bibitem{fritz2012beyond}
Tobias Fritz.
\newblock Beyond bell's theorem: correlation scenarios.
\newblock {\em New Journal of Physics}, 14(10):103001, 2012.

\bibitem{wood2015lesson}
Christopher~J Wood and Robert~W Spekkens.
\newblock The lesson of causal discovery algorithms for quantum correlations:
  Causal explanations of bell-inequality violations require fine-tuning.
\newblock {\em New Journal of Physics}, 17(3):033002, 2015.

\bibitem{Causalbell2015}
R.~Chaves, R.~Kueng, J.~B. Brask, and D.~Gross.
\newblock Unifying framework for relaxations of the causal assumptions in
  bell's theorem.
\newblock {\em Phys. Rev. Lett.}, 114:140403, Apr 2015.
\newblock URL: \url{https://link.aps.org/doi/10.1103/PhysRevLett.114.140403},
  \href {http://dx.doi.org/10.1103/PhysRevLett.114.140403}
  {\path{doi:10.1103/PhysRevLett.114.140403}}.

\bibitem{2019_Orus_NRP_1}
Rom{\'a}n Or{\'u}s.
\newblock Tensor networks for complex quantum systems.
\newblock {\em Nature Reviews Physics}, 1(9):538--550, September 2019.
\newblock \href {http://dx.doi.org/10.1038/s42254-019-0086-7}
  {\path{doi:10.1038/s42254-019-0086-7}}.

\bibitem{2006_Donoho_ITIT_52}
D.~L. Donoho.
\newblock Compressed {{Sensing}}.
\newblock {\em IEEE Trans. Inf. Theor.}, 52(4):1289--1306, April 2006.
\newblock \href {http://dx.doi.org/10.1109/TIT.2006.871582}
  {\path{doi:10.1109/TIT.2006.871582}}.

\bibitem{2007_Candes_PotICoMMA22}
Emmanuel~J. Cand{\`e}s.
\newblock Compressive sampling.
\newblock In {\em Proceedings of the {{International Congress}} of
  {{Mathematicians Madrid}}, {{August}} 22\textendash{}30, 2006}, pages
  1433--1452. May 2007.
\newblock \href {http://dx.doi.org/10.4171/022} {\path{doi:10.4171/022}}.

\bibitem{2012_Eldar}
Yonina~C. Eldar and Gitta Kutyniok.
\newblock {\em Compressed {{Sensing}}: {{Theory}} and {{Applications}}}.
\newblock {Cambridge University Press}, May 2012.

\bibitem{2016_Stern}
Adrian Stern.
\newblock {\em Optical {{Compressive Imaging}}}.
\newblock {CRC Press}, November 2016.

\bibitem{coecke2017picturing}
Bob Coecke and Aleks Kissinger.
\newblock {\em Picturing quantum processes}.
\newblock Cambridge University Press, 2017.

\bibitem{coecke2016categorical}
Bob Coecke and Aleks Kissinger.
\newblock Categorical quantum mechanics ii: Classical-quantum interaction.
\newblock {\em International Journal of Quantum Information}, 14(04):1640020,
  2016.

\bibitem{biamonte2011categorical}
Jacob~D Biamonte, Stephen~R Clark, and Dieter Jaksch.
\newblock Categorical tensor network states.
\newblock {\em AIP Advances}, 1(4):042172, 2011.

\bibitem{wetering2018quantum}
John van~de Wetering.
\newblock Quantum theory is a quasi-stochastic process theory.
\newblock {\em Electronic Proceedings in Theoretical Computer Science}, pages
  179--196, 2018.

\bibitem{fong2013causal}
Brendan Fong.
\newblock Causal theories: A categorical perspective on bayesian networks,
  2013.
\newblock \href {http://arxiv.org/abs/1301.6201} {\path{arXiv:1301.6201}}.

\bibitem{baez2014bayesian}
John~C Baez and Tobias Fritz.
\newblock A bayesian characterization of relative entropy.
\newblock {\em Theory and Applications of Categories}, 29(16):422--456, 2014.

\bibitem{1992_White_PRL_69}
Steven~R. White.
\newblock Density matrix formulation for quantum renormalization groups.
\newblock {\em Physical Review Letters}, 69(19):2863--2866, November 1992.
\newblock \href {http://dx.doi.org/10.1103/PhysRevLett.69.2863}
  {\path{doi:10.1103/PhysRevLett.69.2863}}.

\bibitem{2011_Schollwock_AoP_326}
Ulrich Schollw{\"o}ck.
\newblock The density-matrix renormalization group in the age of matrix product
  states.
\newblock {\em Annals of Physics}, 326(1):96--192, January 2011.
\newblock \href {http://dx.doi.org/10.1016/j.aop.2010.09.012}
  {\path{doi:10.1016/j.aop.2010.09.012}}.

\bibitem{nielsen2015neural}
Michael~A Nielsen.
\newblock {\em Neural networks and deep learning}, volume 2018.
\newblock Determination press San Francisco, CA, USA:, 2015.

\bibitem{abramsky2014operational}
Samson Abramsky and Adam Brandenburger.
\newblock An operational interpretation of negative probabilities and
  no-signalling models.
\newblock In {\em Horizons of the mind. A tribute to Prakash Panangaden}, pages
  59--75. Springer, 2014.

\bibitem{appleby2017introducing}
Marcus Appleby, Christopher~A Fuchs, Blake~C Stacey, and Huangjun Zhu.
\newblock Introducing the qplex: a novel arena for quantum theory.
\newblock {\em The European Physical Journal D}, 71(7):197, 2017.

\bibitem{2011_Becker_SJIS_4}
S.~Becker, J.~Bobin, and E.~Cand{\`e}s.
\newblock {{NESTA}}: {{A Fast}} and {{Accurate First}}-{{Order Method}} for
  {{Sparse Recovery}}.
\newblock {\em SIAM Journal on Imaging Sciences}, 4(1):1--39, January 2011.
\newblock \href {http://dx.doi.org/10.1137/090756855}
  {\path{doi:10.1137/090756855}}.

\bibitem{popescu1994quantum}
Sandu Popescu and Daniel Rohrlich.
\newblock Quantum nonlocality as an axiom.
\newblock {\em Foundations of Physics}, 24(3):379--385, 1994.

\bibitem{penrose1955generalized}
Roger Penrose.
\newblock A generalized inverse for matrices.
\newblock In {\em Mathematical proceedings of the Cambridge philosophical
  society}, volume~51, pages 406--413. Cambridge University Press, 1955.

\bibitem{Spekkens2005}
R.~W. Spekkens.
\newblock Contextuality for preparations, transformations, and unsharp
  measurements.
\newblock {\em Phys. Rev. A}, 71:052108, May 2005.
\newblock URL: \url{https://link.aps.org/doi/10.1103/PhysRevA.71.052108}, \href
  {http://dx.doi.org/10.1103/PhysRevA.71.052108}
  {\path{doi:10.1103/PhysRevA.71.052108}}.

\bibitem{2014_Pfeifer_ACPP_}
{{NCON}}: {{A}} tensor network contractor for {{MATLAB}}.
\newblock February 2014.
\newblock \href {http://arxiv.org/abs/1402.0939} {\path{arXiv:1402.0939}}.

\bibitem{NESTAonline}
{NESTA: A Fast and Accurate First-order Method for Sparse Recovery}.
\newblock \url{https://statweb.stanford.edu/~candes/software/nesta}.

\bibitem{kronmult}
David Gleich.
\newblock Fast and efficient kronecker multiplication.
\newblock
  \url{https://www.mathworks.com/matlabcentral/fileexchange/23606-fast-and-efficient-kronecker-multiplication},
  2020.

\bibitem{Fonseca2015}
E.~A. Fonseca and Fernando Parisio.
\newblock Measure of nonlocality which is maximal for maximally entangled
  qutrits.
\newblock {\em Phys. Rev. A}, 92:030101, Sep 2015.
\newblock URL: \url{https://link.aps.org/doi/10.1103/PhysRevA.92.030101}, \href
  {http://dx.doi.org/10.1103/PhysRevA.92.030101}
  {\path{doi:10.1103/PhysRevA.92.030101}}.

\bibitem{montina2016can}
Alberto Montina and Stefan Wolf.
\newblock Can non-local correlations be discriminated in polynomial time?
\newblock {\em arXiv preprint arXiv:1609.06269}, 2016.

\bibitem{Duarte2018}
Cristhiano Duarte, Samura\'{\i} Brito, Barbara Amaral, and Rafael Chaves.
\newblock Concentration phenomena in the geometry of bell correlations.
\newblock {\em Phys. Rev. A}, 98:062114, Dec 2018.
\newblock URL: \url{https://link.aps.org/doi/10.1103/PhysRevA.98.062114}, \href
  {http://dx.doi.org/10.1103/PhysRevA.98.062114}
  {\path{doi:10.1103/PhysRevA.98.062114}}.

\bibitem{Canabarro2019}
Askery Canabarro, Samura\'{\i} Brito, and Rafael Chaves.
\newblock Machine learning nonlocal correlations.
\newblock {\em Phys. Rev. Lett.}, 122:200401, May 2019.
\newblock URL: \url{https://link.aps.org/doi/10.1103/PhysRevLett.122.200401},
  \href {http://dx.doi.org/10.1103/PhysRevLett.122.200401}
  {\path{doi:10.1103/PhysRevLett.122.200401}}.

\bibitem{wolfe2019quantifying}
Elie Wolfe, David Schmid, Ana~Belén Sainz, Ravi Kunjwal, and Robert~W.
  Spekkens.
\newblock Quantifying bell: the resource theory of nonclassicality of
  common-cause boxes, 2019.
\newblock \href {http://arxiv.org/abs/1903.06311} {\path{arXiv:1903.06311}}.

\bibitem{rosset2019characterizing}
Denis Rosset, David Schmid, and Francesco Buscemi.
\newblock Characterizing nonclassicality of arbitrary distributed devices,
  2019.
\newblock \href {http://arxiv.org/abs/1911.12462} {\path{arXiv:1911.12462}}.

\bibitem{schmid2019typeindependent}
David Schmid, Denis Rosset, and Francesco Buscemi.
\newblock Type-independent resource theory of local operations and shared
  randomness, 2019.
\newblock \href {http://arxiv.org/abs/1909.04065} {\path{arXiv:1909.04065}}.

\bibitem{Navascues2007}
Miguel Navascu\'es, Stefano Pironio, and Antonio Ac\'{\i}n.
\newblock Bounding the set of quantum correlations.
\newblock {\em Phys. Rev. Lett.}, 98:010401, Jan 2007.
\newblock URL: \url{https://link.aps.org/doi/10.1103/PhysRevLett.98.010401},
  \href {http://dx.doi.org/10.1103/PhysRevLett.98.010401}
  {\path{doi:10.1103/PhysRevLett.98.010401}}.

\bibitem{zhu2016quasiprobability}
Huangjun Zhu.
\newblock Quasiprobability representations of quantum mechanics with minimal
  negativity.
\newblock {\em Physical review letters}, 117(12):120404, 2016.

\bibitem{branciard2010characterizing}
Cyril Branciard, Nicolas Gisin, and Stefano Pironio.
\newblock Characterizing the nonlocal correlations created via entanglement
  swapping.
\newblock {\em Physical review letters}, 104(17):170401, 2010.

\bibitem{PhysRevLett.116.010402}
Rafael Chaves.
\newblock Polynomial bell inequalities.
\newblock {\em Phys. Rev. Lett.}, 116:010402, Jan 2016.
\newblock URL: \url{https://link.aps.org/doi/10.1103/PhysRevLett.116.010402},
  \href {http://dx.doi.org/10.1103/PhysRevLett.116.010402}
  {\path{doi:10.1103/PhysRevLett.116.010402}}.

\bibitem{PhysRevLett.116.010403}
Denis Rosset, Cyril Branciard, Tomer~Jack Barnea, Gilles P\"utz, Nicolas
  Brunner, and Nicolas Gisin.
\newblock Nonlinear bell inequalities tailored for quantum networks.
\newblock {\em Phys. Rev. Lett.}, 116:010403, Jan 2016.
\newblock URL: \url{https://link.aps.org/doi/10.1103/PhysRevLett.116.010403},
  \href {http://dx.doi.org/10.1103/PhysRevLett.116.010403}
  {\path{doi:10.1103/PhysRevLett.116.010403}}.

\bibitem{ohlsson2013quadratic}
Henrik Ohlsson, Allen~Y Yang, Roy Dong, Michel Verhaegen, and S~Shankar Sastry.
\newblock Quadratic basis pursuit.
\newblock In {\em Signal Processing with Adaptive Sparse Structured
  Representations (SPARS) Workshop}, 2013.

\end{thebibliography}
\end{document}